\documentstyle[emulateapj]{article} 

\slugcomment{To appear in PASP (2003 Nov)}

\begin{document}

\title{On The Progenitor of the Type II-Plateau Supernova 2003gd \\ in Messier 74\footnote{Based on
observations made with the NASA/ESA {\sl
Hubble Space Telescope}, obtained from the data archive of the Space Telescope
Science Institute, which is operated by the Association of Universities for
Research in Astronomy, Inc., under NASA contract NAS 5-26555.}}

\author{Schuyler D.~Van Dyk}
\affil{IPAC/Caltech, Mailcode 100-22, Pasadena CA  91125}
\authoremail{vandyk@ipac.caltech.edu}

\author{Weidong Li and Alexei V.~Filippenko}
\affil{Department of Astronomy, 601 Campbell Hall, University of
California, Berkeley, CA  94720-3411}
\authoremail{weidong@astro.berkeley.edu, alex@astro.berkeley.edu}

\begin{abstract}
{\sl Hubble Space Telescope\/} ({\sl HST}) WFPC2 archival F606W and F300W
images obtained within one year prior to the explosion of the nearby Type II
supernova (SN) 2003gd in Messier 74 (NGC 628) have been analyzed to isolate the
progenitor star.  The SN site was located using precise astrometry applied to
the {\sl HST\/} images.  Two plausible candidates are identified within
$0{\farcs}6$ of the SN position in the F606W image.  Neither
candidate was detected in the F300W image.  SN 2003gd appears to be
of Type II-plateau (II-P), with age $\approx$ 87~d on June 17 UT and with low 
reddening [$E(B-V) = 0.13$ mag].  The
most likely of the two progenitor candidates has $M_{V_0} \approx -3.5$ mag
(for an extinction-corrected distance modulus of 29.3 mag) and, 
based on additional color 
information derived from a high-quality,
archival ground-based $I$-band image, we estimate that this star was a red
supergiant with initial (zero-age main sequence)
mass $M_{\rm ZAMS} \approx 8$ -- 9 $M_{\odot}$.  This mass
estimate is somewhat lower than, but relatively consistent with, recent limits 
placed on the progenitor masses of other SNe II-P, using {\sl HST\/} data.  
Future {\sl HST\/} imaging with the {\sl HST\/}
Advanced Camera for Surveys, when the SN has faded considerably, will be
very useful in pinpointing the exact SN location and securing
identification of the progenitor. If our proposed candidate is confirmed,
it will be only the sixth SN progenitor ever directly identified.
\end{abstract}


\section{Introduction}

Determining which stars give rise to supernovae (SNe) is at the heart of SN
research. The main obstacle is that a SN leaves few traces of the
progenitor star.  Only a small handful of historical SNe (SN 1961V [Zwicky
1964, 1965], SN 1978K [Ryder et al.~1993], SN 1987A [e.g., Gilmozzi et
al.~1987; Sonneborn, Altner, \& Kirshner 1987], SN 1993J [Aldering, Humphreys,
\& Richmond 1994; Cohen, Darling, \& Porter 1995], and SN 1997bs [Van Dyk et
al. 1999, 2000]) have had precursors identified. Additionally, it
should be noted that these five SNe were all at least somewhat unusual.  Van
Dyk, Li, \& Filippenko (2003a) recently attempted the direct identification of
the progenitors of 16 relatively normal Type II and Type Ib/c SNe, using
archival images from the {\sl Hubble Space Telescope\/} ({\sl HST}).  They may
have identified the progenitors of the Type II SNe 1999br and 1999ev, the Type
Ib SNe 2001B and 2001is, and the Type Ic SN 1999bu.  Subsequent follow-up to
the tentative identification of the progenitor of the Type II SN 2001du, using
an {\sl HST\/} image of the SN itself, resulted in an upper limit to the
progenitor's mass (Van Dyk, Li, \& Filippenko 2003b; see also Smartt et
al.~2003a).

Here we attempt to identify the progenitor of a recent and relatively nearby
Type II-plateau (II-P) event.  SN 2003gd was visually discovered by Evans
(2003) on June 12.82 UT at about 13.2 mag, and about 20\arcsec\ east and
150\arcsec\ south of the nucleus of Messier 74 (M74; NGC 628).  McNaught
(2003a) confirmed the discovery on CCD images and measured an accurate optical
position for the SN of $\alpha$(J2000) = $1^h36^m42{\fs}65$, $\delta$(J2000) =
$+15\arcdeg 44\arcmin 19{\farcs}9$ (McNaught 2003b).  Garnavich \& Bass (2003a)
obtained near-infrared (0.85--2.4 $\mu$m) spectra of SN 2003gd on June 13.46
UT; the strong, broad Paschen-line emission in the $J$-band order of the
spectrum led them to classify the SN as Type II.  This classification has since
been confirmed by Kotak, Meikle, \& Smartt (2003) and by Phillips (2003), both
groups suggesting that the SN was discovered at an age of $\sim$1--2 months.
From ground-based images, Garnavich \& Bass (2003b) place a limit on the
progenitor star brightness of $R > 24$ mag at four months before discovery.

M74 was also the host to the peculiar Type Ic SN 2002ap (e.g., Mazzali et
al.~2002; Leonard et al. 2002c; Foley et al.~2003).  Both Sharina, Karachentsev,
\& Tikhonov (1996) and Sohn \& Davidge (1996) measured a distance to M74, based
on photometry of the brightest stars in the galaxy.  Their consistent results
indicate a true (extinction-corrected) distance modulus $\mu_0 = 29.3$ mag, 
corresponding to a distance of about 7.2 Mpc.  Lacking any additional distance 
information, we assume this distance throughout this paper.

\section{Analysis}

Pre-SN archival {\sl HST\/} WFPC2 images containing the SN site were obtained
by program GO-9676, using several different pointings, on 2002 August 25 and 28
UT in the bands F606W and F300W, for total exposure times of 3100 s and 3000 s,
respectively.  A summary of the available {\sl HST\/} data is given in Table 1.
Recent multi-band images of SN 2002ap were obtained by GO-9114 in 2003 January
using the Advanced Camera for Surveys (ACS), but, unfortunately for us, these
were made with the High-Resolution Camera (HRC).  The HRC's field-of-view is
too small, and the displacement of SN 2002ap from SN 2003gd is too large, for
these images to be useful in detecting the progenitor. (Furthermore, at the
time of this writing, these images were still proprietary.)

It is absolutely essential to locate the SN site in the {\sl HST\/}
images with high astrometric precision.
From a $V$-band image obtained with the Las Campanas Observatory 1.0-m
telescope and kindly provided by M.~Hamuy, we have measured a precise position
for the SN of $\alpha$(J2000.0) = $1^h36^m42{\fs}67$, $\delta$(J2000.0) = $+15\arcdeg
44\arcmin 19{\farcs}7$, with a total uncertainty of $0{\farcs}2$, using the 
Two Micron All Sky Survey (2MASS) as the astrometric reference.  Note the
excellent agreement with the position measured by NcNaught (2003b); we adopt
the average of these two measurements for the SN position, i.e., $\alpha$(J2000.0) =
$1^h36^m42{\fs}66$, $\delta$(J2000.0) = $+15\arcdeg 44\arcmin 19{\farcs}8$.  
Position data for the SN are summarized in Table 2.  

Applying the astrometric method described by Van Dyk et al.~(2003a) to a {\it
wmosaic\/} of the {\sl HST\/} F606W coadded image pair, U8IXCY02M and
U8IXCY03M, utilizing a deep $V$-band image of M74 obtained at the Palomar 1.5-m
telescope (see Foley et al.~2003) and again adopting 2MASS as a reference, we are
able to isolate the position of the SN on the mosaic to $\pm 0{\farcs}6$.
This uncertainty is the SN absolute position uncertainty ($0{\farcs}2$), the
uncertainty in the astrometric grid applied to the Palomar image ($0{\farcs}4$),
and the uncertainty in the grid further applied to the {\sl HST\/} mosaic
($0{\farcs}4$), as well as the relative difference between the absolute position
measurements ($0{\farcs}1$) and the relative accuracy of the 2MASS astrometry
($\lesssim 0{\farcs}1$), all added in quadrature.  The SN site is located on
the WF2 chip, and the position uncertainty, at $0{\farcs}1$ pixel$^{-1}$ for
the WF chips, therefore corresponds to $\pm$6 WF pixels.

Since the various pointings are
difficult to properly align, both within and between the two bands, we elected
to apply the photometry routine HSTphot (Dolphin 2000a, 2000b) to the images in each
band in separate units, and to subsequently combine the results.  (In fact, we
were unable to derive the necessary offsets between the two F300W exposures
U8IXCA04M and U8IXCA05M, and the third, U8IXCA03M, as a result of the low
signal-to-noise ratio in all of these images.)  HSTphot automatically accounts
for WFPC2 point-spread function (PSF) variations and charge-transfer effects
across the chips, zeropoints, aperture corrections, etc.  In this case the
HSTphot output was in the flight magnitude system.
In Figure 1 we show the SN environment in the F606W band.  Two sources, A and
B, are detected by HSTphot very near or within the error circle.  

Table 3 lists the photometry in the F606W band for Stars A and B.  The F606W
magnitudes are the uncertainty-weighted average of the photometry performed on
the U8IXCY01M + U8IXCY02M + U8IXCY03M image trio and on the U8IXCA01M +
U8IXCA02M image pair.  
Unfortunately, the F300W exposures were not sensitive enough, since 
Stars A and B were not detected in the F300W images, to $m_{\rm F300W} 
\gtrsim 23.8$ mag (3$\sigma$).

Two other stars are seen in Figure 1 about $0{\farcs}5$ 
to the northeast of the error circle edge; the 
easternmost one has $m_{\rm F606W}=25.09 \pm 0.08$ mag, and the westernmost
one has $m_{\rm F606W}=25.79 \pm 0.23$ mag (it appears to be somewhat blended).
Another two stars are seen near Star A, one $\sim 0{\farcs}4$ due east, 
and one $\sim 0{\farcs}3$ due
south; the eastern one has $m_{\rm F606W}=25.97 \pm 0.12$ mag, and the
southern one has $m_{\rm F606W}=26.02 \pm 0.16$ mag.
Several fainter
sources are also seen within the error circle in Figure 1, but they are 
undetected by HSTphot.

\section{Discussion}

We can attempt to estimate the masses of Stars A and B and therefore comment on
the plausibility of each as the likely SN progenitor.  For this we need an
estimate for the stars' color, as well as the reddening toward the SN, and the
metallicity of the SN environment.  From the F300W images, limits on the color
$m_{\rm F300W}-m_{\rm F606W}$ are $\gtrsim -1.6$ mag for Star A and $\gtrsim
-2.8$ for Star B.  Using the transformations from flight system to
Johnson-Cousins magnitudes via SYNPHOT, applied to the Bruzual Spectral
Synthetic Atlas (see Van Dyk, Filippenko, \& Li 2002), this translates to $U-V
\gtrsim -0.9$ mag for Star A.  The flight system color for Star B is too blue
to realistically transform to a standard color, but it is likely $U-V >
-1.5$ mag.

Additional color information for both progenitor candidates can be obtained
from a high-quality $I$-band image (seeing $\sim 0{\farcs}9$) obtained with 
the 2.6-m Nordic Optical Telescope (NOT; see Larsen \& Richtler 1999), 
archived and made available online by NED\footnote{NED is the
NASA/IPAC Extragalactic Database, http://nedwww.ipac.caltech.edu.}.
Figure 2 shows the SN environment, which is approximately the same field shown
in Figure 1.  Matching the $I$-band and F606W images was not trivial, 
even after geometrically transforming one image relative to the other,
due to the differences in pixel scale and resolution.  However,
two faint objects are seen in Figure 2 very near the positions of both
Stars A and B (the identification of the counterpart to Star B on the $I$-band 
image is less certain than that for Star A).  
Assuming these are the $I$-band counterparts to Stars A and B, this implies 
that these two stars
are red.  Using PSF-fitting photometry in DAOPHOT/ALLSTAR 
(Stetson 1987, 1992) within
IRAF\footnote{IRAF (Image Reduction and Analysis Facility) is distributed by
the National Optical Astronomy Observatories, which are operated by the
Association of Universities for Research in Astronomy, Inc., under cooperative
agreement with the National Science Foundation.} and, calibrating this
image using the comparison stars in Foley et al.~(2003), we find the $I$-band
magnitudes for the two stars listed in Table 3.  Applying the 
transformations derived from SYNPHOT we convert $m_{\rm F606W}-I$ for both
stars to $V-I$, and $m_{\rm F606W}$ to $V$ (for such red stars, $V$ is
$\sim 0.8$ mag fainter than $m_{\rm F606W}$); these values, and their estimated
uncertainties, are also listed in Table 3.

We have obtained $BVRI$ images of SN 2003gd on a number of epochs with the 
Katzman Automatic Imaging Telescope (KAIT; see Li et al.~2000; Filippenko et 
al.~2001).  Again, we have calibrated these images using the comparison stars 
for SN 2002ap in Foley et al.~(2003).  The SN photometry is listed in Table 4.
The SN was already quite red at the time of discovery.
Unfortunately, the SN color is subject to a degeneracy between the SN age and
the reddening.  We can try to break this degeneracy through a comparison with
the very well-studied Type II-P SN 1999em in NGC 1637 (Hamuy et 
al.~2001; see also Leonard et al.~2002a and Elmhamdi et al.~2003).  

In Figure 3 we show the $BVRI$ light curves of SN 2003gd and for comparison
the light curves of SN 1999em, adjusted to the true distance modulus of M74.
No reddening correction is made to the light curve of SN 1999em to match 
that of SN 2003gd.
(However, we have adjusted the light curves of both SNe in time to find the 
best match.)  
Similarly, in Figure 4 we show the color evolution of SN 2003gd, and
for comparison, that of SN 1999em, with both SNe corrected for reddening 
[$E(B-V) = 0.1$ mag for SN 1999em, Leonard et al. 2002c; $E(B-V) = 0.13$ mag
for SN 2003gd, see below].  
Both figures imply rather convincingly that SN 2003gd is also a SN II-P and 
that it was quite old at the time of discovery (near the 
end of the plateau phase of the light curves).  Similar to SN 1999em, 
SN 2003gd suffers relatively low reddening, which is
also supported
by the lack of interstellar Na I D absorption, as well as low continuum 
polarization seen in spectra of SN 2003gd we have obtained at the Lick 3-m 
and Keck 10-m telescopes.  
Although the photometry for the two SNe II-P agrees quite well
during the plateau phase,
it is most striking how much fainter SN 2003gd is 
relative to SN 1999em in the late-time nebular phase.  The faint nebular-phase 
tail and large post-plateau drop may be suggestive of low ejected $^{56}$Ni 
mass, similar to the subluminous SNe II-P SN 1997D and 1999br (Benetti et 
al.~2001; Zampieri et al.~2003).  A more detailed study of SN 2003gd is 
clearly warranted.

From the plateau we estimate an age for SN 2003gd of $87 \pm 3$ d on 
June 17 UT, placing the date of explosion at about March 17 UT, consistent
with the range of explosion dates between February and April suggested by
Garnavich \& Bass (2003b).  Additionally, 
we estimate the total reddening to SN 2003gd as $E(B-V) = 0.13 \pm 0.03$ mag 
[the Galactic reddening toward M74 is itself quite low, $E(B-V) =
0.07$ mag; Schlegel, Finkbeiner, \& Davis 1998; and NED].

In Figure 5 we show the absolute $V$ magnitude and $U-V$ color limits from 
the {\sl HST\/} photometry for the two progenitor candidates, corrected for 
reddening to the SN, assuming the Cardelli, Clayton, \& Mathis (1989) 
reddening law, and adjusted for the true M74 distance modulus.  SN 2003gd 
occurred about 161\arcsec\ southeast of the M74 nucleus, or at 
$R/R_{25} \approx 0.5$.  At this nuclear distance, van Zee et al.~(1998) 
represent the relative oxygen abundance as log(O/H) + 12 $\approx$ 9 dex.  The 
metallicity in the SN 2003gd environment,
then, is possibly $\sim$1.5 times greater than solar (where the solar O/H
abundance is 8.8 dex; Grevesse \& Sauval 1998).  In Figure 5 we therefore show
for comparison the model stellar evolutionary tracks for a range of masses from
Lejeune \& Schaerer (2001), assuming enhanced mass loss for the most massive
stars and a metallicity $Z = 0.04$ for the SN environment.

From the positions of Stars A and B in Figure 5, we cannot rule out that the SN
progenitor was a blue star.  However, except for the most massive stars, which
may evolve back to the blue toward the end of their lives (such as seen for the
40 $M_{\odot}$ model), we do not expect SN~II progenitors to generally occupy
this color region [$(U-V)_0 \approx -1$ to $-2$ mag], since lower-mass stars
are still on (or are just barely off) the main sequence and therefore not
sufficiently evolved.  Also, we expect SN~II-P progenitors to be red
supergiants, since both the optical P-Cygni spectral line profiles and the
plateau phase of the SN light curve (arising from a hydrogen recombination wave
in the envelope) require such extensive hydrogen envelopes.  Furthermore, the
lack of radio emission from SN 2003gd (C. J. Stockdale et al. 2003, private
communication) implies a paucity of circumstellar matter, providing additional evidence
that most of the progenitor's matter was still contained in the star itself at
the time of explosion.

In Figure 6 we show the absolute magnitude and color for Stars A and B from the
combined {\sl HST\/} and NOT photometry, again corrected for reddening to the
SN and adjusted for the true M74 distance modulus.  For comparison we again 
show the Lejeune \& Schaerer (2001) evolutionary tracks, as in Figure 5 (for 
$9\ M_{\odot}$ we also show the track corresponding to solar metallicity,
$Z = 0.02$; as can be seen, little difference exists in the red between the 
tracks of different metallicity).  From Figure 6
it appears that both stars are red supergiants, 
and that Star A had an initial (zero-age main sequence) 
mass $M_{\rm ZAMS} \approx \ 8$--9 $M_{\odot}$.  The fainter Star B 
appears to have
had an initial mass $M_{\rm ZAMS} \approx 5\ M_{\odot}$, which is formally
below the theoretical lower limit for core-collapse SNe ($\sim 8\ M_{\odot}$;
e.g., Woosley \& Weaver 1986).

One of the two stars is most likely the progenitor: the F606W detection limit,
$m_{\rm F606W}=27.3$ mag (3$\sigma$), when corrected for the true M74 distance 
modulus and assuming stars of the same reddening-corrected $V-I$ as Stars A and
B, corresponds to $M_{V_0} \approx -1.6$ mag and therefore to initial masses 
well below the core-collapse limit.  Thus,
we can probably rule out that the progenitor is not detected in the F606W image.
Although Star A is farther from the SN position than Star B, and just outside 
the edge
of the error circle, we consider Star A to be the most likely progenitor
candidate, based on its estimated initial mass.  From Figure 2 it also
appears to be the most plausible candidate, being the brightest object in 
the $I$ band within the SN's larger, $\sim$1{\arcsec} radius, environment.  
(Although of comparable brightness to Stars A and B in F606W, the 
four other stars mentioned in \S~2 which are 
outside, but near, the error circle
do not have clearly identifiable counterparts on the $I$-band image, implying
that these stars are bluer than Stars A and B and can probably be discounted 
as progenitor candidates.)  The estimate for the SN
2003dg progenitor mass is lower than the limits derived for the nearby SNe
II-P 1999gi ($15^{+5}_{-3}\ M_{\odot}$; Leonard et al.~2002b) and 1999em ($20 \pm
5\ M_{\odot}$; Leonard et al.~2003), but is consistent with that for the SN II-P 2001du 
($13^{+7}_{-4}\ M_{\odot}$; Van Dyk et al.~2003b).  Our estimates for the absolute
brightness and initial mass for the SN 2003gd progenitor are consistent with those
reported by Smartt, Maund, \& Hendry (2003b).

\newpage

\section{Conclusions}

Examining {\sl HST\/} archival WFPC2 images of the Type II-P SN 2003gd in M74
obtained before explosion, we have isolated two possible SN progenitor stars.
Although we cannot exclude a blue SN progenitor, it is more likely that the
progenitor was red.  This is further supported by the brightness of the two
stars on a ground-based archival $I$-band image.  SN 2003gd is
an old SN II-P, and we have estimated that its reddening is
quite low, $E(B-V) = 0.13$ mag, similar to that of the well-studied SN
II-P 1999em.  We estimate that the more likely candidate of the two stars had
an initial mass $M_{\rm ZAMS} \approx 8$--9 $M_{\odot}$ which, together with
mass limits derived for other SNe~II-P, suggests that such SNe II-P arise from
massive stars at the lower extreme of the possible mass range for core collapse.

It would be most fruitful to recover the SN in multiple bands with {\sl HST\/}
at high spatial resolution, preferably with ACS, 
when the SN has significantly dimmed.  This will allow us to pinpoint the SN's
exact location on the preexplosion images and thus be definitive about the
progenitor's identification. If our proposed candidate is confirmed, it will be
only the sixth SN progenitor ever directly identified. Additionally, the
multiband imaging of the stars in the environment of the fading SN would
provide possible further constraints on the age and mass of the progenitor
star, based on the characteristics of its surviving neighbors.

\acknowledgements

We thank the referee, Sydney van den Bergh, 
for useful comments that improved the paper.  We also
thank Andrew Dolphin for assistance with HSTphot, and Doug Leonard for
illuminating discussions regarding the 
photometric evolution of SN 2003gd and other SNe II-P.
This publication makes use of data products from the Two Micron All Sky Survey,
which is a joint project of the University of Massachusetts and the
IPAC/California Institute of Technology, funded by NASA and NSF.  This research
also utilizes the NASA/IPAC Extragalactic Database (NED) which is
operated by the Jet Propulsion Laboratory, California Institute of Technology,
under contract with NASA.  The work of A.V.F.'s group at UC Berkeley is
supported by NSF grant AST-0307894, as well as by 
NASA grants AR-9953, AR-9529, and AR-8754 from the Space Telescope
Science Institute, which is operated by AURA, Inc., under NASA contract
NAS5-26555. KAIT was made possible by generous donations from Sun Microsystems,
Inc., the Hewlett-Packard Company, AutoScope Corporation, Lick Observatory, the
National Science Foundation, the University of California, and the Sylvia and
Jim Katzman Foundation.

\clearpage

\begin{deluxetable}{lccl}
\def\phmm{\phm{$-$}}
\tablenum{1}
\tablecolumns{4}
\tablecaption{Summary of {\sl HST\/} Observations\tablenotemark{a}}
\tablehead{\colhead{Date} & \colhead{Filter}
& \colhead{Exp.~Time} & \colhead{{\sl HST}} \nl
\colhead{(UT)} & \colhead{} & \colhead{(s)} & \colhead{Dataset}}
\startdata
2002 Aug 25 & F606W &  500 & U8IXCA01M \nl
            &       &      & U8IXCA02M \nl
            & F300W & 1000 & U8IXCA03M \nl
            &       &      & U8IXCA04M \nl
            &       &      & U8IXCA05M \nl
2002 Aug 28 & F606W &  700 & U8IXCY01M \nl
            &       &      & U8IXCY02M \nl
            &       &      & U8IXCY03M \nl
\enddata
\tablenotetext{a}{Program GO-9676.}
\end{deluxetable}


\begin{deluxetable}{cccc}
\tablenum{2}
\tablewidth{4.5truein}
\tablecolumns{4}
\tablecaption{Position Data for SN 2003gd}
\tablehead{
\colhead{$\alpha$(J2000)} & \colhead{$\delta$(J2000)} & \colhead{Uncertainty}& \colhead{Source}}
\startdata
$1^h36^m42{\fs}65$ & $+15\arcdeg 44\arcmin 19{\farcs}9$ & $0{\farcs}1$\tablenotemark{a} & McNaught (2003b) \\
$1^h36^m42{\fs}67$ & $+15\arcdeg 44\arcmin 19{\farcs}7$ & $0{\farcs}2$\tablenotemark{b} & This paper \\
$1^h36^m42{\fs}66$ & $+15\arcdeg 44\arcmin 19{\farcs}8$ & $0{\farcs}2$\tablenotemark{b} & Average \\
\enddata
\tablenotetext{a}{Uncertainty in each coordinate.}
\tablenotetext{b}{Total uncertainty.}
\end{deluxetable}


\begin{deluxetable}{ccccc}
\tablenum{3}
\tablewidth{4.2truein}
\tablecolumns{5}
\tablecaption{Photometry of Stars Near SN 2003gd (mag)}
\tablehead{
\colhead{Star} & \colhead{$m_{\rm F606W}$} & \colhead{$I$} & \colhead{$V-I$} & 
\colhead{$V$}}
\startdata
A & $25.39 \pm 0.09$ & $22.5 \pm 0.2$ & $3.6 \pm 0.3$ & $26.2 \pm 0.2$ \nl
B & $26.56 \pm 0.24$ & $23.8 \pm 0.4$ & $3.5 \pm 0.4$ & $27.4 \pm 0.3$ \nl
\enddata
\end{deluxetable}

\clearpage

\begin{deluxetable}{cccccc}
\tablenum{4}
\tablewidth{5.5truein}
\tablecolumns{5}
\tablecaption{KAIT Photometry of SN 2003gd (mag)}
\tablehead{
\colhead{UT date} &\colhead{JD--2450000 } & \colhead{$B$} & \colhead{ $V$ } &
\colhead{$R$} & \colhead{$I$}
}
\startdata
Jun 17 & 2807.99 & 15.44(04) & 14.14(04) & 13.65(03) & 13.26(03)\\
Jun 20 & 2810.98 & 15.54(02) & 14.19(02) & 13.71(02) & 13.32(02)\\
Jun 22 & 2812.97 & 15.55(02) & 14.24(02) & 13.72(03) & 13.34(02)\\
Jun 27 & 2817.96 & 15.70(02) & 14.33(02) & 13.81(03) & 13.43(02)\\
Jun 29 & 2819.97 & 15.74(02) & 14.37(03) & 13.82(02) & 13.42(03)\\
Jul 3  & 2823.95 & 15.88(02) & 14.46(02) & 13.89(02) & 13.50(02)\\
Jul 5  & 2825.97 & 15.94(02) & 14.49(02) & 13.93(02) & 13.52(02)\\
Jul 7  & 2827.95 & 16.03(02) & 14.56(02) & 13.97(02) & 13.56(02)\\
Jul 9  & 2829.96 & 16.15(02) & 14.65(02) & 14.05(02) & 13.64(02)\\
Jul 11 & 2831.95 & 16.25(02) & 14.74(03) & 14.15(02) & 13.75(03)\\
Jul 13 & 2833.95 & 16.35(05) & 14.87(03) & 14.24(03) & 13.81(02)\\
Jul 15 & 2835.95 & 16.62(04) & 15.07(03) & 14.42(03) & 13.98(03)\\
Jul 17 & 2837.97 & 16.91(02) & 15.33(02) & 14.62(02) & 14.17(03)\\
Jul 23 & 2843.96 & 18.31(02) & 16.85(02) & 15.94(03) & 15.41(04)\\
Jul 26 & 2846.97 & 18.69(03) & 17.32(03) & 16.26(03) & 15.69(03)\\
Jul 29 & 2849.97 & 18.94(05) & 17.42(02) & 16.40(02) & 15.81(03)\\
Aug 4  & 2855.97 & 19.05(05) & 17.45(03) & 16.34(03) & 15.81(04)\\
Aug 10 & 2861.97 & 19.07(06) & 17.56(03) & 16.47(02) & 15.89(03)\\
Aug 19 & 2870.97 & 19.14(07) & 17.66(03) & 16.56(02) & 16.00(03)\\
\enddata
\tablenotetext{}{Note: uncertainties (hundredths of a magnitude) are 
indicated in parentheses.}
\end{deluxetable}


\begin{figure}
\figurenum{1}
\caption{The site of SN 2003gd in the archival 3100-s F606W composite image
from 2002 late-August.  The error circle has radius $0{\farcs}6$.  Two
progenitor candidates, Stars A and B, near or within the circle are indicated 
with short line segments.}
\end{figure}


\begin{figure}
\figurenum{2}
\caption{The environment of SN 2003gd, as seen in an
$I$-band image of the host galaxy, M74, obtained with the 2.6-m 
Nordic Optical Telescope (see Larsen \& Richtler 1999).  Two objects are
seen in the image at the approximate locations of Stars A and B, and are
labelled accordingly in the figure.}
\end{figure}

\clearpage

\begin{figure}
\figurenum{3}
\plotfiddle{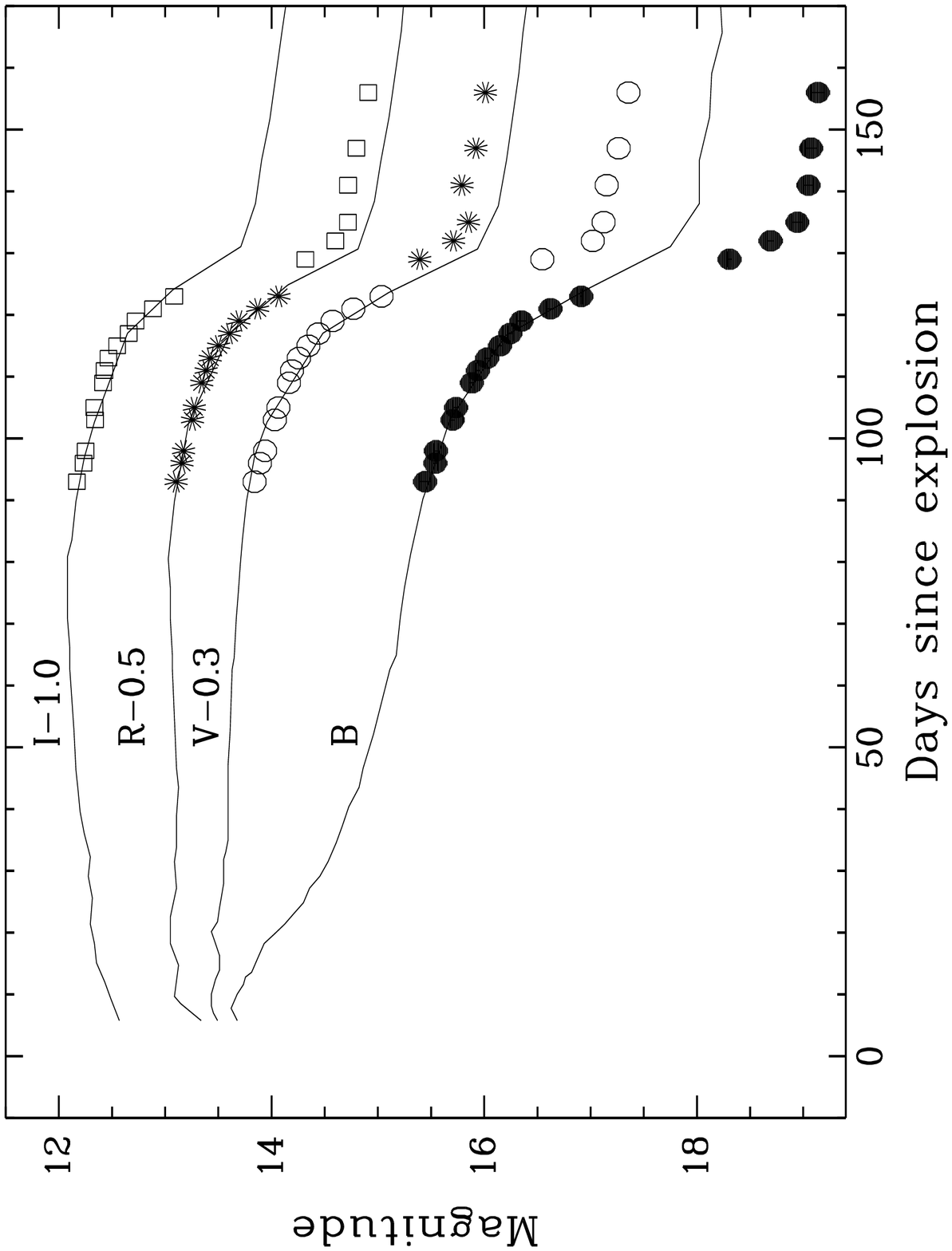}{300pt}{-90}{65}{65}{-250}{+400}
\caption{The $BVRI$ light curves for SN 2003gd from KAIT observations.
For comparison, the light curves for the well-studied Type II-plateau
SN 1999em (Hamuy et al.~2001) are shown, 
adjusted to the true distance modulus of M74.  No additional reddening 
correction has been applied to the SN 1999em light curves.}
\end{figure}

\clearpage

\begin{figure}
\figurenum{4}
\plotfiddle{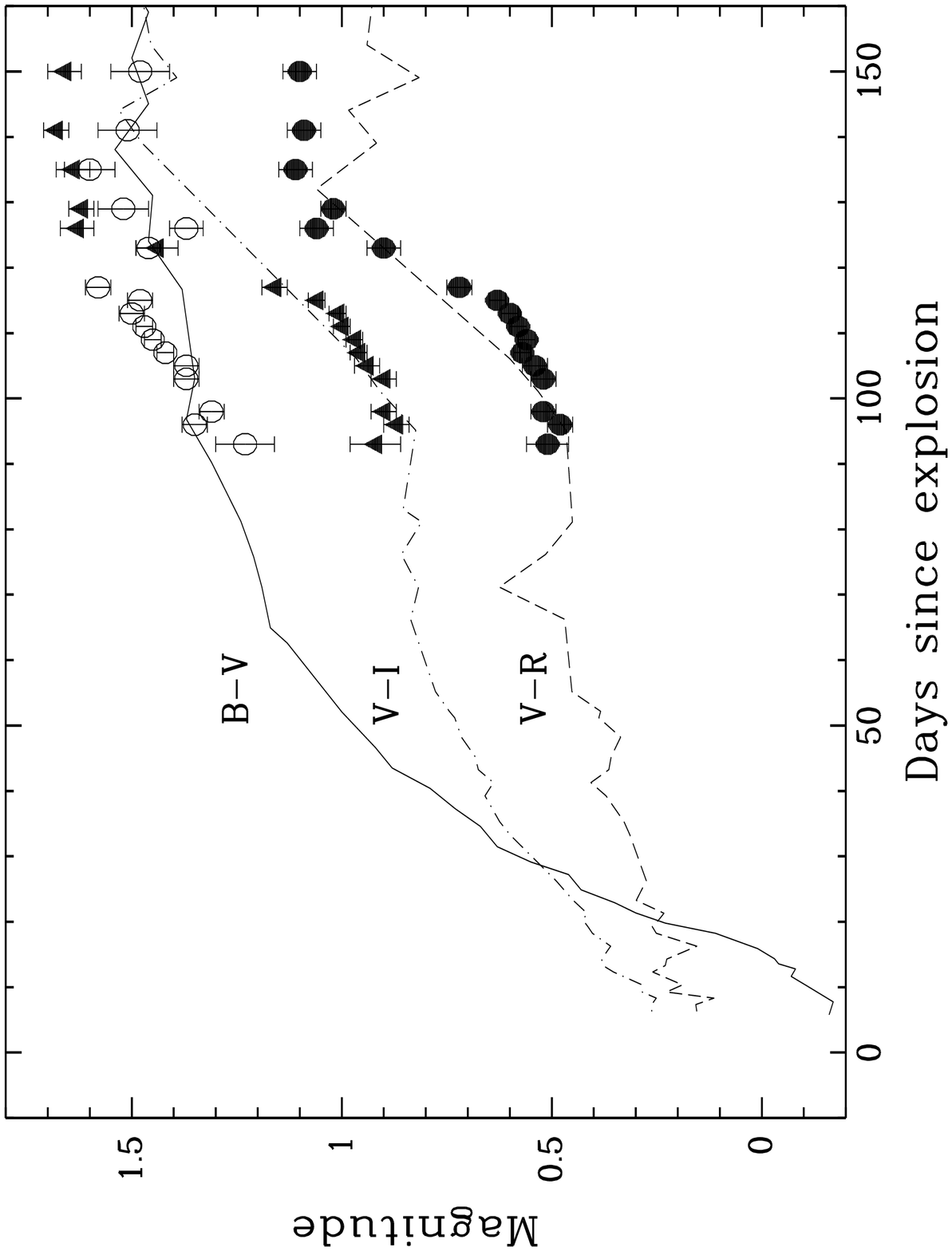}{300pt}{-90}{65}{65}{-250}{+400}
\caption{The $B-V$ ({\it open circles}), $V-R$ ({\it filled circles}), 
and $V-I$ ({\it stars}) colors of SN 2003gd from KAIT observations, corrected
for an assumed reddening $E(B-V) = 0.13$ mag.  
For comparison, the reddening-corrected color evolution for SN 1999em [Hamuy et
al.~2001; $E(B-V) = 0.1$ mag] is shown for
$B-V$ ({\it solid line}), $V-R$ ({\it dashed line}), and $V-I$
({\it dot-dashed line}).}
\end{figure}

\clearpage

\begin{figure}
\figurenum{5}
\plotone{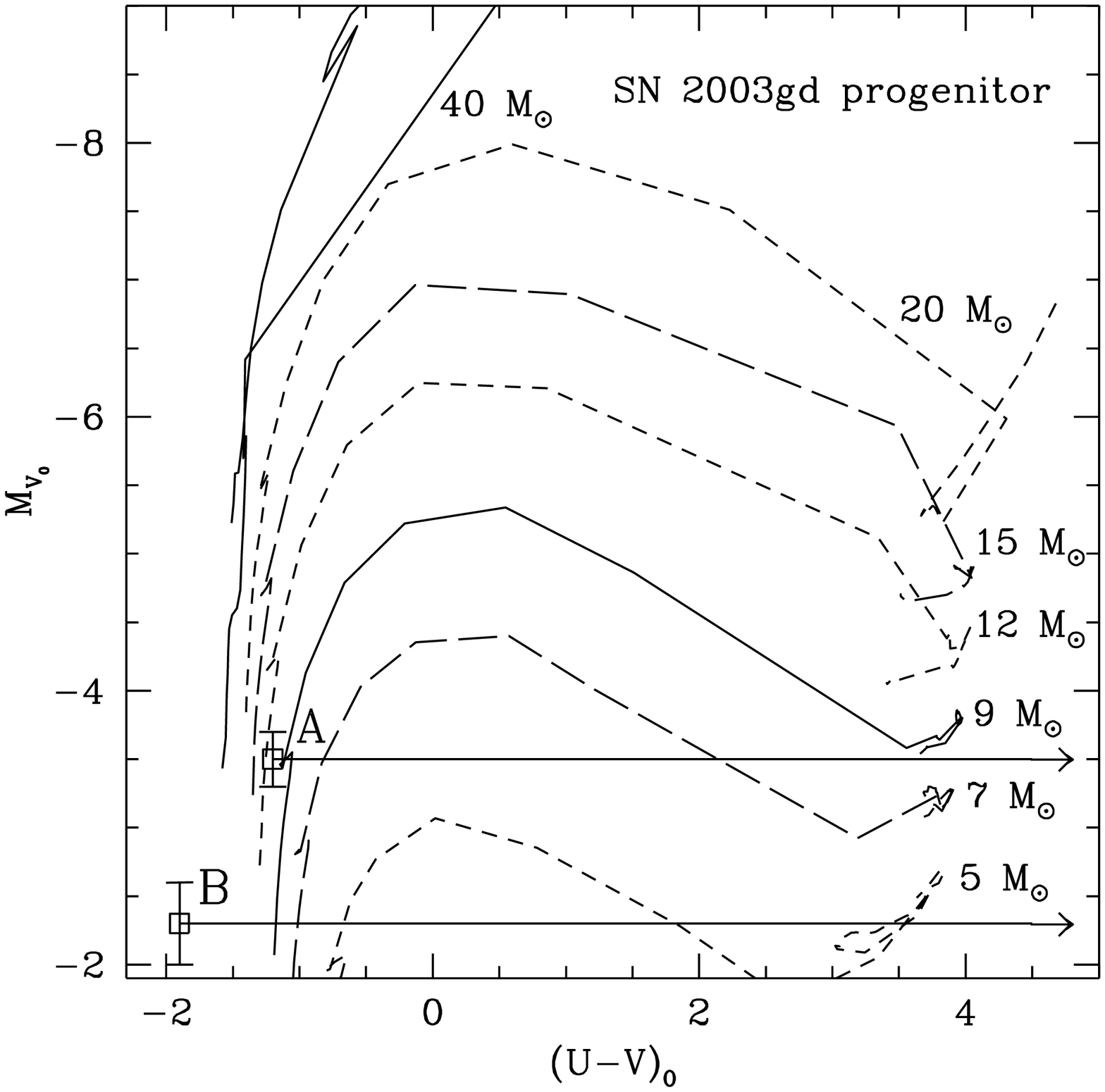}
\caption{Intrinsic color-magnitude diagram showing model stellar evolutionary
tracks (alternating {\it long-dashed lines\/}, {\it short-dashed lines\/}, and
{\it solid lines}) for a range of masses from Lejeune \& Schaerer (2001), with
enhanced mass loss for the most massive stars and a metallicity $Z = 0.04$.
Also shown are the absolute extinction-free magnitudes and reddening-corrected
color limits for the two progenitor candidates, Stars A and B, assuming $E(B-V)
= 0.13$ mag and an extinction-corrected distance modulus to M74 of $\mu_0 = 29.3$ 
mag.} 
\end{figure}

\clearpage

\begin{figure}
\figurenum{6}
\plotone{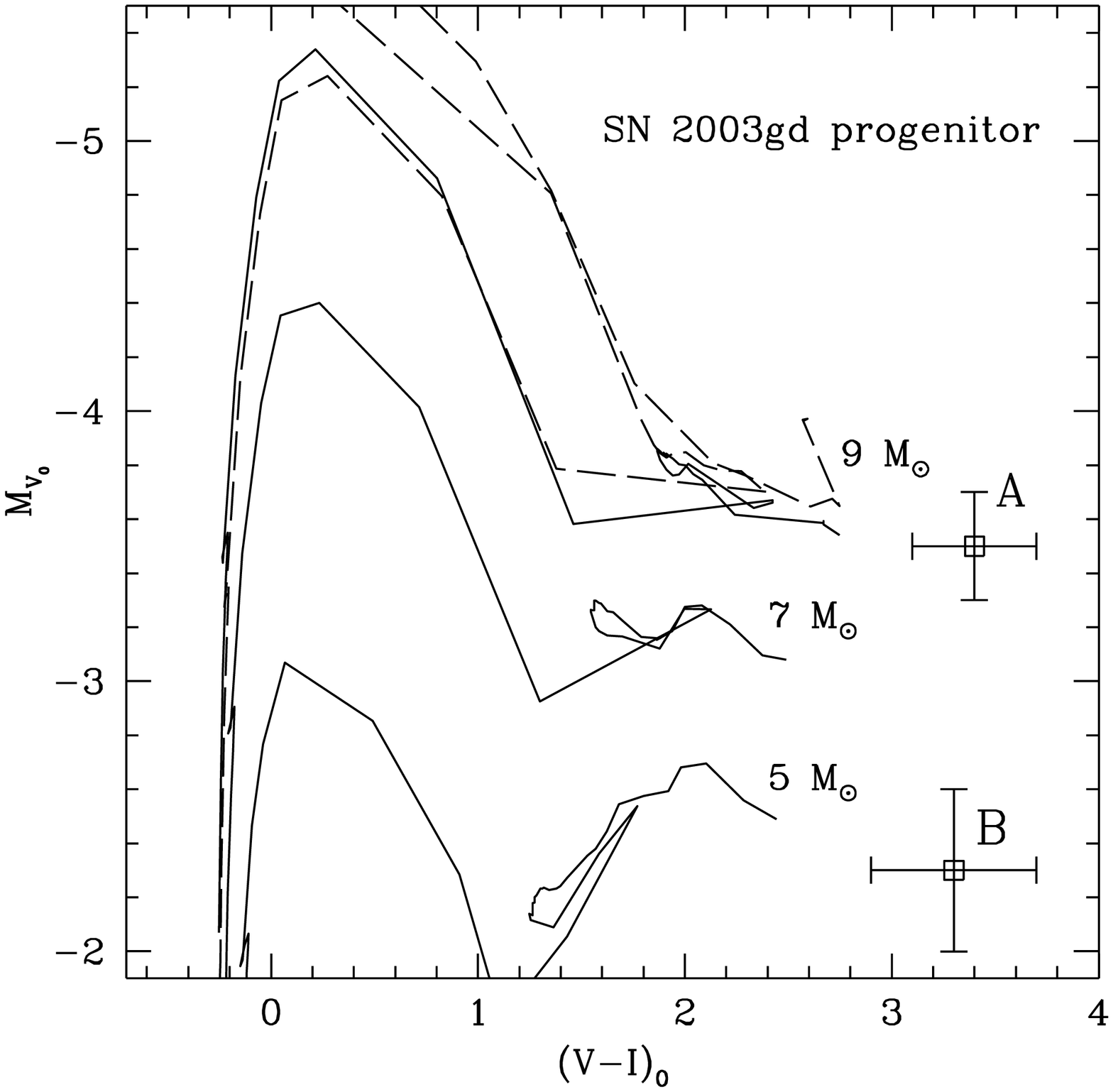}
\caption{Intrinsic color-magnitude diagram showing model stellar evolutionary
tracks for a range of masses from Lejeune \& Schaerer (2001), with
enhanced mass loss for the most massive stars and a metallicity $Z = 0.04$
({\it solid lines}), and the track for $9\ M_{\odot}$ and metallicity 
$Z = 0.02$ ({\it dashed line}).
Also shown are the absolute extinction-free magnitudes and reddening-corrected
colors for the two progenitor candidates, Stars A and B, assuming $E(B-V) =
0.13$ mag and an extinction-corrected distance modulus to M74 of $\mu_0 = 29.3$
mag.}
\end{figure}

\end{document}